\documentclass[preprint]{aastex}

\slugcomment{Accepted, ApJ Letters.}

\shorttitle{Convective Zone Masses}
\shortauthors{Pinsonneault et al.}

\begin{document}


\title{The Mass of the Convective Zone in FGK Main Sequence Stars and
the Effect of Accreted Planetary Material on Apparent Metallicity
Determinations}


\author{M. H. Pinsonneault, D. L. DePoy, and M. Coffee}
\affil{Department of Astronomy, Ohio State University}



\begin{abstract}
The mass of the outer convective zone in FGK main sequence stars decreases
dramatically with stellar mass. Therefore, any contamination
of a star's atmosphere by accreted planetary material should affect
hotter stars much more than cool stars. If recent suggestions that 
high metal abundances in stars
with planets are caused by planetesimal accretion are correct, 
then metallicity enhancements in earlier-type stars with planets
should be very pronounced. No such trend is seen, however.   
\end{abstract}




\section{Introduction}

There are now approximately fifty planets known around 
other stars (Marcy et al. 2000).
This is a sufficiently large number that
statistically meaningful trends among the properties
of the stars-with-planets (SWPs)
sample may help understand aspects of planet formation, evolution, and
distribution. For example, there are several claims
that SWPs have higher metallicities than stars of similar spectral type in
the solar neighborhood  (e.g. Gonzalez 1997). 
The metallicities of FGK dwarfs in the solar
neighborhood are strongly peaked around {\rm [Fe/H]}$\approx -0.2$; less
than 25\%
of these stars have {\rm [Fe/H]} $>$ 0.0 and less than 5\% 
have {\rm [Fe/H]} $>$ 0.2 (see Wyse \& Gilmore 1995 and 
Rocha-Pinto \& Machiel 1998).
This result is based largely on medium-band photometry of a large number of
stars (Twarog 1980; Olsen 1983; Olsen 1993), although it is consistent
with the smaller sample (91 G \& K dwarfs) observed spectroscopically
by Favata et al. (1997). 
In contrast, Gonzalez \& Laws (2000) and
Santos et al. (2000) use high resolution
spectroscopic observations to show that SWPs are comparatively 
metal rich: $\sim$90\%
have {\rm [Fe/H]}$>$0.0 and $\sim$60\% have {\rm [Fe/H]}$>$0.2.
Similar results are reported by
Gonzalez (1998), Gonzalez \& Vanture (1998), 
and Gonzalez, Wallerstein, \& Saar (1999).
Note that Gimenez (2000) shows that
medium-band photometry of SWPs also gives consistenly high {\rm [Fe/H]}
values. Formally, the
statistical significance of this difference between the
metallicity of SWPs and FGK dwarfs in the solar neighborhood 
is very high.
However, there is some possibility that systematic effects could
affect the reality of the difference; e.g. that some observational
selection effect allows planets around metal rich stars to be detected
with greater ease (although see Butler et al. 2000).

Suppose, however, that SWPs are indeed apparently more metal 
rich than other stars. One popular
explanation is that the apparent increase in the observed metallicity is due
to accretion of planetary material, which enriches the star's convective
envelope 
(Lin et al. 1996; Rasio \& Ford 1996; Murray et al. 1998; 
Quillen \& Holman 2000).
For example,
the dissolution of Jupiter into the Sun's convective envelope would raise the
observed metallicity of the Sun by roughly 0.09 dex (assuming Jupiter is
roughly 10\% metals in solar proportions). 
The stability of the convective envelope throughout the main
sequence life of a star suggests that this elevated metallicity would
be observed for very long times, assuming the accretion took place
after the star arrived on the main sequence (Laughlin \& Adams 1997).
Murray et al. (2000) examine a related possibility: the accretion of
$\sim$0.5 M$_\earth$ of rocky material during the heavy bombardment
phase.

In this paper we examine the possibility that accretion of planetary material
could appreciably affect the observed metallicities of stars. We find that
hotter, more massive stars should be dramatically more sensitive to accretion, 
but that this is contrary to what is observed. We suspect, therefore, that 
accretion is not the primary cause of the enhanced metallicities in SWPs.

\section{Models}

We constructed stellar models in the effective temperature range 
where giant planets have been observed; this corresponds to masses of
0.6-1.3 in 0.1 solar mass increments.  These models did not 
include the effects of rotational mixing or microscopic diffusion
(used in this paper to include the effects of gravitational 
settling, thermal diffusion, and radiative levitation);
the possible impact of these phenomena is discussed later.  The basic 
physics in these standard models has been described elsewhere (Sills,
Pinsonneault, \& Terndrup 2000); we summarize the basic ingredients here.

Our base case is constructed with the solar heavy element to hydrogen
ratio Z/X of 0.0245 from the heavy element mixture of Grevesse \& Noels
(1993).  We use the OPAL equation of 
state (Rogers, Swenson, \& Iglesias 1996) 
where available and the Saumon, Chabrier, \& van Horn (1995)
equation of state in regions of parameter space where OPAL data is
not available; we ramp between the two EOS at the edges of the OPAL
tables.  Nuclear reaction rates are taken from Adelberger et al. (1998);
following the results of Gruzinov \& Bahcall  (1998) we adopt the
approximation of weak screening for nuclear reaction rates.
We use OPAL95 opacities (Iglesias \& Rogers 1996) above $10^4$ K and
molecular opacities from Alexander \& Ferguson (1994) below.  For
the solar abundance models we use Allard \& Hauschildt (1995) 
model atmospheres
at $\tau = \frac {2}{3}$ for a surface boundary condition.
We calibrate our models to reproduce the solar radius
($6.9598 \times 10^{10}$ cm) and luminosity ($3.844 \times 10^{33}
\frac {erg}{s}$) at the solar age (4.57 Gyr; see Bahcall \&
Pinsonneault 1995.)  This gave a solar Y = 0.266 and mixing length
$\alpha = 1.74$.  Our models were started in the pre-MS at the mass-radius
relationship given by Stahler (1988).  These models were evolved to an 
age of 4.57 Gyr to test for the dependence of convection zone depth on age.

The Allard \& Hauschildt (1995) atmospheres are not available for a 
fine grid of metal abundances near solar, so we constructed ZAMS
models with [Fe/H] from -0.3 to +0.2 in 0.1 dex increments using 
the Kurucz (1994) model atmospheres as a surface boundary conditions.  For
the purposes of this paper, the choice of surface boundary condition does not have a major impact on the predicted depth of the convection zone as a function of effective temperature.

Figure 1 shows the predicted mass of the convective zone as a function of stellar effective temperature. Large changes in the
assumed intial metallicity and age of the star make very little
difference to the result; the maximum difference over a wide set of
input initial conditions is only $\sim$0.15 dex.  There is a 
straightforward physical explanation: the depth of convection zones 
in main sequence stars depends primarily on the position within the 
star of the hydrogen and helium ionization zones.  It is a well-known 
result of stellar structure and evolution that along the main 
sequence the temperature decreases and the density increases at fixed 
mass fraction as the total stellar mass decreases.  This can be 
demonstrated analytically by simple homology relationship (cf. Kippenhahn 
\& Thomas 1991).  These effects cause ionization zones where the 
opacity is high to be systematically located in deeper layers for lower 
mass stars.  There will therefore be progressively deeper surface convection 
zones for lower mass stars.  The effective temperature is a useful 
proxy for this: stars will necessarily have low density in the 
photosphere, so the effective temperature of the star sets the 
starting point in the $\rho - T$ plane which determines the depth of 
ionization zones.  We note that stars evolving off the main sequence 
obey a similar, if not identical, convection zone depth as a function 
of effective temperature; the principal physical difference is the lower 
surface gravity in these stars.  Our result also implies that the 
surface convection zone depth of stars with enhanced metal abundances 
only in the convective envelope (through enrichment by planetesimals, 
for example) will be the similar to those of normal stars of the same 
effective temperature.

The models demonstrate
that the stellar effective temperature is the best 
diagnostic of the depth of the surface convection zone.
The effective temperature of a star is also directly observable.
This correspondence between a theoretically meaningful metric
and an observable quantity is fortunate and eliminates the need
to compute or estimate quantities such as stellar mass and age.

\section{The Metallicities of Stars with Planets}

Figure 2 shows {\rm [Fe/H]} versus stellar effective
temperature (T$_{eff}$) for 33 SWPs derived from high
resolution spectra by
Gonzalez et al. (2000), Santos et al. (2000), Gonzalez et al. (1999),
Gonzalez \& Vanture (1998), and Gonzalez \& Laws (2000).
There may be a slight trend of increasing metallicity with increasing
T$_{eff}$ as noted by Laughlin (2000)
and Gonzalez et al. (2001). Also shown in Figure 2 is the effect of adding 
1 $M_\earth$, 3 $M_\earth$, and 10 $M_\earth$ of Fe to 
the convection zone of stars with a range of initial {\rm [Fe/H]}. 
This is roughly the amount of
Fe present in 0.5 $M_{Jup}$, 1.7 $M_{Jup}$, and 5.5 $M_{Jup}$ planets
(see Murray et al. 2001 and Guillot 1999).

Roughly half of the relatively cool SWPs (T$_{eff}$ $<$ 5800K) have very high
values of metallicities: {\rm [Fe/H]} $>$ 0.3. If these stars began their
main sequence lives with {\rm [Fe/H]} $\sim$ $-$0.2 to 0.0, then they
must have accreted at least $\sim$10 $M_\earth$ of Fe. 
If the hotter stars had
accreted this amount of Fe, however, the {\rm [Fe/H]} should be $>$ 0.6
for any star with T$_{eff}$ $>$ 6000K. Clearly, this pattern is not
observed. In fact, there are no stars known 
with {\rm [Fe/H]} $>$ 0.6 (Castro et al. 1997). Further, the two most 
metal rich ({\rm [Fe/H]} $\approx$ 0.5)
stars in the SWP sample have T$_{eff}$ $\approx$ 5300K. If these
stars began with {\rm [Fe/H]} $\approx$ 0.0, they must have 
accreted $\sim$25 $M_\earth$ of Fe. A star with T$_{eff}$ $\approx$ 6200K
that accreted this amount of Fe would have {\rm [Fe/H]} $\approx$ 1!
Presumably, such an extremely unusual star would already been identified
in the solar neighborhood.

Alternatively, if the hotter stars
intially had {\rm [Fe/H]} $\sim$ $-$0.2 to 0.0, then they could have
accreted no more than $\sim$1 $M_\earth$ of Fe. 
If the cooler stars accreted this amount of Fe, then they must have
started with {\rm [Fe/H]} $\sim$ $+$0.2, substantially higher than the
solar neighborhood mean.

\section{Impact of Non-standard Processes}

In standard stellar models convection is the only mixing mechanism, so 
it is straightforward to convert a given contribution of heavy elements 
onto a star into an enhancement in its surface [Fe/H].  There are three 
major classes of effects that could dilute enhancements of the heavy 
elements in surface layers of stars: mass loss, microscopic diffusion, 
and envelope mixing driven by stellar rotation. We find that none of 
these has a major impact on our conclusions. Mass loss, for example,
removes part of the total envelope mass 
of a star, so a surface metal enhancement is
reduced for two reasons: there is
a lower total 
mass and T$_{eff}$ (therefore a deeper surface CZ and more dilution) and 
enriched material is preferentially removed.
The first effect is taken care of by considering the current T$_{eff}$.
The second effect is small because main sequence mass loss 
rates are very small.  The 
solar mass loss rate is of order $10^{-14}$ solar masses per year; this 
implies a total mass loss over the solar lifetime of order $5 \times 10^{-5}$ 
solar masses.  Even if this much lost matter is replaced with undiluted 
material, there is an insignificant 
correction to the 
total surface abundance even in the hottest stars we consider.

Diffusion is the process whereby
heavy elements will sink with respect to light ones 
in the presence of a gravitational potential; this can be counteracted 
by radiative levitation for partially ionized species.  The latter effect 
is most important for stars with much thinner surface convection zones than 
we consider (see Richer, Michaud, \& Turcotte 2000 for an example). 
Modern calculations provide estimates of the rate of 
gravitational settling/thermal 
diffusion that significantly improve the agreement between helioseismic data 
and solar models (e.g. Bahcall, Pinsonneault, \& Basu 2001).  The overall 
effect in solar models is small, roughly 10\% for heavy 
elements by the age of the Sun.
The effect would increase to higher masses and effective 
temperatures, reaching $\sim$30\% at 6400K.
The surface helium abundance would drop by the same process, boosting
the reduction in the surface {\rm [Fe/H]} by a modest amount.

The convection zone depth changes by a factor of 
twenty across the mass range of SWPs, so gravitational settling 
corresponds to a perturbation on a strong mass-dependent trend.  
Furthermore, the difference in the degree of metal diffusion is 
partially counteracted by the longer main sequence lifetimes of the 
lower mass stars; solar mass stars would experience approximately 
a 20\% effect by the end of their main sequence evolution.  
Also, diffusion may be inhibited by other processes in stars
hotter than the Sun.
Lithium data can also be used to constrain mass-dependent diffusion 
processes (Chaboyer et al. 1992).  Lithium settles at approximately 
the same rate as helium; both settle faster than heavier elements.  
The absence of a strong downwards trend in halo star lithium abundances 
with increased effective temperature sets constraints on the 
mass dependence of element segregation in these stars.  
Because they are older and hotter than the range we are considering, 
this in turn implies that such effects will be even smaller 
in the systems with planets. 
Strong effects are possible in hotter stars in the vicinity of 
the mid-F star lithium dip (Boesgaard \& Trippicco 1986;
see however Deliyannis et al. 1998), 
but planets have not yet 
been detected in this mass range.  We conclude that the inclusion of 
metal diffusion may have a modest ($\sim$10-15\%) effect on the 
surface metal abundances of stars (which 
affects the zero-point of the dilution), 
but will not have a strong impact on the predicted behavior of metal 
overabundance as a function of mass.

There is strong empirical evidence for mild envelope mixing 
driven by rotation in lower 
main sequence stars (see Pinsonneault 1997).
Lithium is a useful diagnostic since it is destroyed 
around 2.5$\times$10$^6$ K, so any mixing process that reaches 
this temperature can affect the surface abundance.  Standard models predict 
lithium depletion consistent
with observations in young open clusters (e.g. Soderblom et al. 1993).  
However, measured lithium abundances show a marked decline 
on the main sequence.
The meteoritic lithium abundance {\rm [Li/H]} is 3.3 (where
H = 12). Young cluster stars have {\rm [Li/H]} $\sim$ 3.2 for stars
with T$_{eff}$ above 6000K, declining to $\sim$2.0 at 5600K and $\sim$2.8
at 5000K. By comparison, the solar photospheric {\rm [Li/H]}
is 1.1, while typical abundances at 6200K in M67 
are 2.2-2.6 (Jones, Fischer, \& Soderblom 1999).
Relative to young main sequence solar 
analogues, the Sun is depleted by roughly 1.7 dex, while stars 
in the old open cluster M67 at a temperature of 6200 K are depleted by 
about 0.6-1.0 dex relative to the meteoritic abundance.  This 
phenomenon could therefore possibly have significant impact 
on surface metal enrichment from the infall of 
planetesimals.

However, there are some important differences between the mixing of 
heavy elements that do not experience nuclear processing and lithium, 
which is destroyed.  Solar models which include both heavy element diffusion 
and mixing sufficient to explain lithium remove only 
about 25\% of the effects of metal diffusion (Richard et al. 1996; 
Bahcall, Pinsonneault, \& Basu 2001).
Metal diffusion creates a composition gradient at the base of the
surface convective zone during the main sequence, which is
similar to what would occur with planetesimal accretion.
The difference between the 
lithium depletion for the Sun and the lithium depletion for the 
hotter stars is largely a reflection of the greater distance 
between the base of the surface convection zone and the temperature where 
lithium is destroyed.

To estimate the interaction of mixing and envelope metal enhancement, we 
constructed two simplified models with 1.0 and 1.2 M$_\odot$ and ran 
them from the pre-MS to an age of 100 Myr.  We used an 
initial {\rm [Fe/H]} of +0.2; the convection zone depths 
were of order 0.032 and 0.006 M$_\odot$.
respectively.  We then enforced solid body rotation at a period of 2.5 days 
(typical for young open cluster stars) and artificially enhanced the 
surface abundance of a trace element by a factor of 10 to mimic the sudden 
increase in metal abundance from accretion.  The model was then evolved 
with rotational mixing and angular momentum loss as described by 
Pinsonneault et al. (1999) and gravitational settling of metals and heavy 
elements (Bahcall \& Pinsonneault 1995.)  Our primary interest was the 
relationship between surface lithium depletion and the reduction of any
surface metallicity enhancement.

As expected, there was an initial transient in the surface abundance from 
the presence of a step function in abundance of the trace element and the 
time delay for lithium-depleted material to reach the surface.  We 
therefore believe that the most reliable technique is to compare the 
late-time behavior of the depletion; from stellar lithium data the degree 
of mixing between 100 Myr and 500 Myr is roughly equivalent to the degree 
of mixing between 500 Myr and 5 Gyr because of the strong decline in 
stellar rotation with increased age.

Between 500 Myr and 4 Gyr, we found a net reduction of 0.35 dex in Li and 
0.22 dex in the surface heavy 
element abundances in the hotter star ($\sim$6050K); 
the cooler star ($\sim$5500 K) had a net decrease of 0.55 dex in Li and 0.1 
dex in the heavy element abundance.  These rough estimates 
would be doubled
from the zero-age main sequence, if the depletion factor at early 
epochs was similar to the late-time behavior.  Starting from 100 Myr 
abundances of 3.1 and 2.7 respectively, this would imply Hyades-age surface 
lithium abundances of 2.8 and 2.15, with M67 age abundances of 2.45 and 1.6 
respectively.  These are consistent with open cluster observations. This is 
across a 500 K range and a factor of five in surface convection zone 
depth.  By comparison, the SWPs at 6100 K have surface lithium abundances 
in the range of 2.4-2.7, which would correspond to a 0.2-0.4 dex decrease 
of a large 1 dex initial enhancement of the surface metal abundance.

An even more significant clue as to the limited impact of mixing is the 
star Tau Boo (HD 120136), which has a low surface lithium abundance of 1.0, 
a temperature of 6400 K, and an [Fe/H] of +0.32 (Gonzalez \& Laws 
2000).  The low lithium abundance is a sign of extensive mixing and yet 
the surface abundance is still very high.  If SWPs were lower metallicity 
objects with polluted surfaces, we would expect this star to have {\it 
low} metal abundance unless the initial abundance was absurdly 
high ({\rm [Fe/H]} $\sim$ 2!).  
Furthermore, the lithium abundances of SWPs do not appear to be 
drastically different from other stars of comparable mass and age (Ryan 
2000), which further suggests that they have not experienced unusual 
depletion patterns.  We therefore believe that mixing in the temperature 
range below the lithium dip will not have a dramatic impact on our 
conclusions. A separate paper with more detailed discussion of these
effects is in preparation.

\section{Conclusions}

Careful modeling shows that the convective zone masses in FGK main 
sequence stars decreases dramatically with stellar mass. 
Convective zone masses in F stars, for example, are more 
than 10 times less massive than in K stars. 
Thus, we expect that accretion of
planetary material by stars while on the main sequence should lead to
extremely high apparent metallicities in early-type stars.
Physical processes not included in standard models, such as
rotational mixing and diffusion, will provide only modest
changes in this trend.
In particular, the measured surface lithium abundances for some SWPs
constrains the degree to which extra mixing could have reduced any 
initial enrichment. The models demonstrate
that the measured stellar effective temperature is a good diagnostic 
of the surface convection zone depth across a wide range of metallicities 
and main sequence ages.

Data obtained to date on SWPs suggests that there is
no strong trend of metallicity enhancement
with stellar effective temperature.  Therefore, we conclude that
unless SWPs somehow know to accrete planetary material in a manner
that is inversely proportional to their own stellar mass (and
in a very non-linear way), then it seems unlikely that the 
possibly enhanced metallicites of SWPs is caused by accretion of
planetary material.

If the suggestion that SWPs are more metal-rich than typical stars
is confirmed, our results imply that SWPs have high surface metal abundances 
because they were born that way, rather than requiring significant 
pollution of their outer layers by accretion of planetary material.  
This suggests a metallicity 
threshold for the formation of (at least) massive planets close 
to stars.  We cannot 
exclude modest infall which would not cause a large surface metallicity 
enhancement (at the $<$ 1 $M_\earth$ level) or models where doomed
Jupiters penetrate the surface convective zone (Sandquist et al. 1998).  
If pollution of the outer layers of 
stars by infall is significant, the best sample in which to 
test this idea is in stars with T$_{eff}$ around 6200K, 
since the surface convective zones in these stars have little
mass and even a small amount of accretion would result in very high
apparent metal abundances.



\acknowledgments

This work was stimulated by discussions during the morning coffee of the OSU
Astronomy Department. These daily doses of caffeine and discussion involve many
members of the department, but R. Pogge, D. Weinberg, and D. Terndrup were
particularly vocal on this subject.




\clearpage


\figcaption{
Convective Zone Mass versus Effective Temperature for main
sequence stars. The thin lines show the effect of metallicity ({\rm [Fe/H]}
varies from -0.3 to +0.2); the thick lines the effect of age (from 0.1 to
4.57 Gyr). The physics input into the two sets of models is also slightly
different (see text). The maximum difference in the models
is $\sim$0.15 dex. 
}

\figcaption{
T$_{eff}$ and {\rm [Fe/H]} derived from high resolution spectra
for 33 SWPs (see text for references).
The uncertainties in T$_{eff}$ are typically estimated to be 25-50K;
the uncertainties in the {\rm [Fe/H]} shown are as given in the references.
The effect of adding 1 $M_\earth$, 3 $M_\earth$, and 10 $M_\earth$ 
masses of Fe 
to a main sequence star and contaminating the star's convective zone is shown
by lines in the individual panels. In each panel the stars are assumed to
have a different initial {\rm [Fe/H]} as noted.
The lines do not represent a good fit to the data points. Note
that no star is known with ${\rm [Fe/H]} > +0.6$.
}

\begin{figure}
\plotone{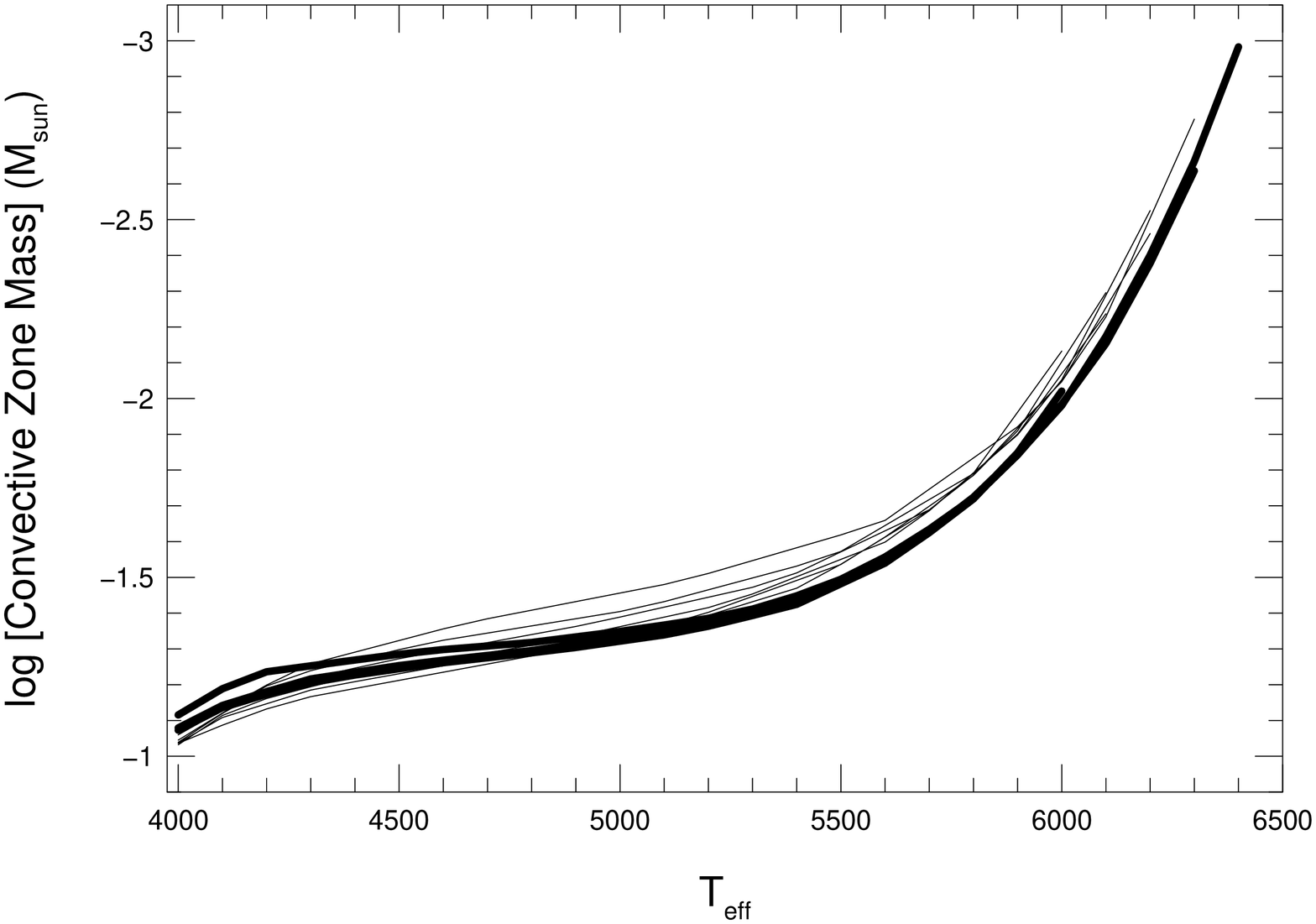}
\end{figure}

\begin{figure}
\plotone{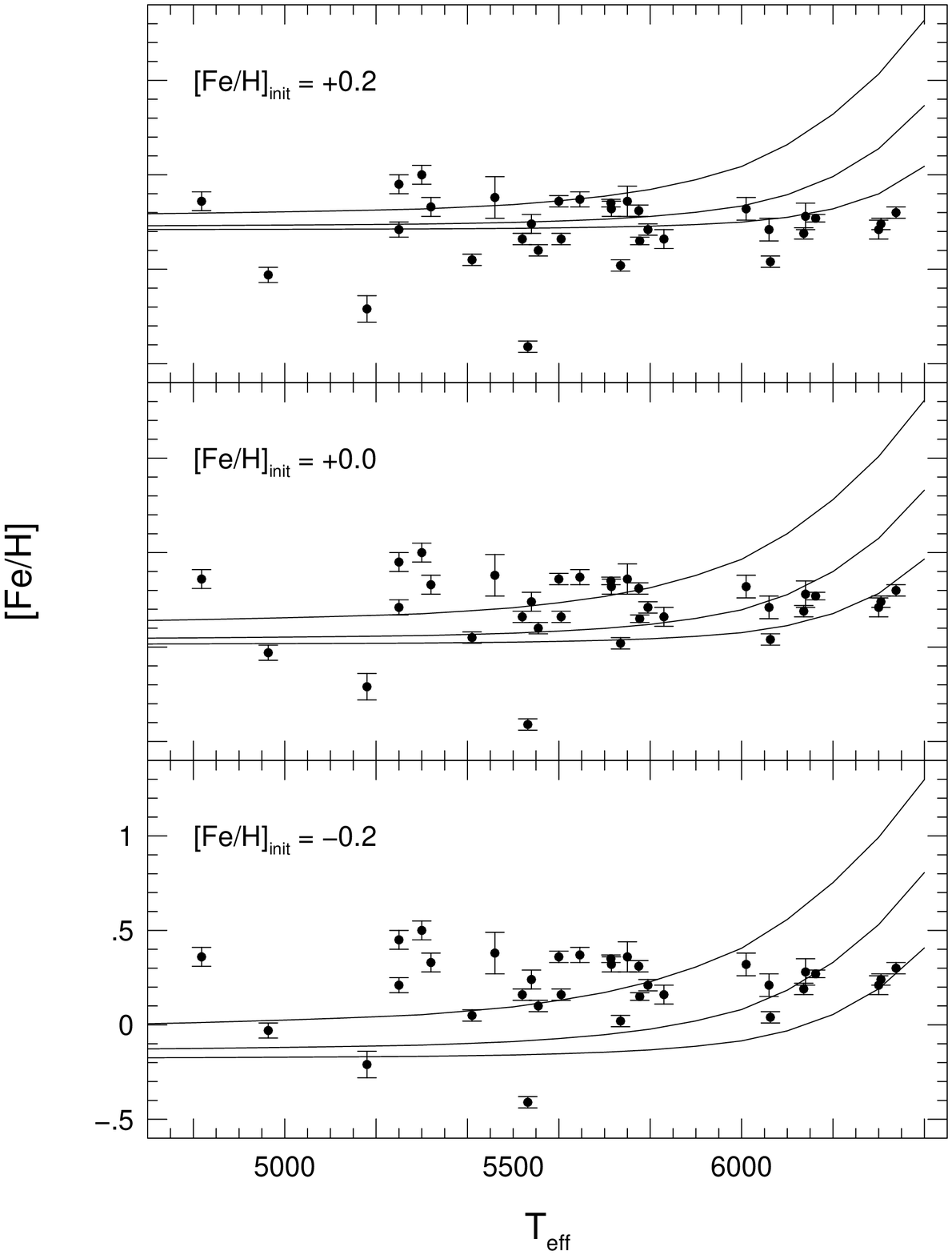}
\end{figure}


\begin{thebibliography}{}
\bibitem{a98}Adelberger, E.C. et al. 1998, Rev.Mod.Phys., 70, 1265
\bibitem{a99}Alexander, D.R. \& Ferguson, J.W. 1994, ApJ, 437, 879
\bibitem{a97}Allard, F. \& Hauschildt, P.H. 1995, ApJ, 445, 433
\bibitem{b95}Bahcall, J.N. \& Pinsonneault, M.H. 1995, 
Rev. Mod. Phys. 67, 781
\bibitem{bpb01}Bahcall, J.N., Pinsonneault, M.H., \& Basu, S. 
2001, ApJ, in press
\bibitem{bt86}Boesgaard, A.M. \& Tripicco, M.J. 1986, ApJLett, 302, L49
\bibitem{b00}Butler, R. P., Vogt, S., Marcy, G. W., Fischer, D. A.,
Henry, G. W., \& Apps, K. 2000, ApJ, 545, 504
\bibitem{c97}Castro, S.,
Rich, M. R., Grenon, M., Barbuy, B., \& McCarthy, J. K.
1997, AJ, 114, 376.
\bibitem{deli}Deliyannis, C.P., Boesgaard, A.M., 
Stephens, A., King, J.R., Vogt, S.S. \& 
Keane, M.J. 1998, ApJLett, 498, L147
\bibitem{f97}Favata, F. Micela, G., \& Sciortino, S. 1997, A\&A, 323, 809
\bibitem{g00}Gimenez, A. 2000, A\&A, 356, 213
\bibitem{g97}Gonzalez, G. 1997, MNRAS, 285, 403.
\bibitem{g98}Gonzalez, G. 1998, A\&A, 334, 221.
\bibitem{gl00}Gonzalez, G., \& Laws, C. 
2000, AJ, 119, 390.
\bibitem{gonzalez2001} Gonzalez, G., Laws, C., Tyagi,
S., \& Reddy, B. E. 2001, AJ, 121, 432
\bibitem{gv98}Gonzalez, G., \& Vanture, A. D. 1998,
A\&A, 339, L29.
\bibitem{gws99}Gonzalez, G., 
Wallerstein, G., \& Saar, S. 1999, ApJ, 511, L111.
\bibitem{gn}Grevess, N. \& Noels, A. 1993, in 
Origin and Evolution of the Elements, ed. 
N Prantos, E. Vangione-Flam, \& M. Casse (Cambridge: Cambridge University 
Press), 15
\bibitem{gb98}Gruzinov, A. \& Bahcall, J.N. 1998, ApJ, ApJ, 504, 996
\bibitem{gt99}Guillot, T. 1999, Plan. Sp. Sci., 47, 1183.
\bibitem{i96}Iglesias, C. \& Rogers, F. 1996, ApJ, 464, 943
\bibitem{j99}Jones, B.F., Fischer, D. \& Soderblom, 
D.R. 1999, AJ, 117, 330
\bibitem{k91}Kippenhahn, R. \& Weigart, A. 1991, 
Stellar Structure and Evolution, (New York: Springer-Verlag)
\bibitem{k94}Kurucz, R. 1994, Solar abundance model 
atmospheres for 0,1,2,4,8 km/s. 
\bibitem{k95}Kurucz CD-ROM No. 19. 
Cambridge, Mass.: Smithsonian Astrophysical Observatory.
\bibitem{l00}Laughlin, G. 2000, submitted to ApJ.
\bibitem{la97}Laughlin, G., \& Adams, F. C. 1997,
ApJ, 491, L51.
\bibitem{lbr96}Lin, D. N. C.,
Bodenheimer, P., \& Richardson, D. C. 1996, Nature, 380, 606.
\bibitem{marcy2000} Marcy, G. W., Cochran, W. D., \& Mayor,
M. 2000, in Protostars and Planets IV, ed. V. Mannings, A. P. Boss, \& S. S.
Russell (Tucson:University of Arizona Press), p. 1285.
\bibitem{m98}Murray, N., Hansen, B., Holman, M., \&
Tremaine, S. 1998, Science, 279, 69.
\bibitem{m00}Murray, N., Chaboyer, B., Hansen, B., 
Arras, P. \& Noyes, B. 2000, astro-ph/0011530
\bibitem{o83}Olsen, E. H. 1983, A\&AS, 54, 55.
\bibitem{o93}Olsen, E. H. 1993, A\&AS, 102, 89.
\bibitem{p97}Pinsonneault, M.H. 1997, ARAA, 35, 557
\bibitem{p99}Pinsonneault, M.H., Walker, T., 
Steigman, G., \& Narayanan, V.K. 1999, ApJ, 527, 180
\bibitem{qh00}Quillen, A. C., \& Holman, M. 2000,
AJ, 119, 397.
\bibitem{rf96}Rasio, F. A., \& Ford, E. B. 1996,
Science, 274, 954.
\bibitem{r96}Richard, O., Vauclair, S., 
Charbonnel, C. \& Dziembowski, W.A. 1996, Astr.Ap., 312, 1000
\bibitem{ryan}Ryan, S.G. 2000, M.N.R.A.S., 316, L35
\bibitem{r97}Richer, J., Michaud, G. \& Turcotte, S. 2000, ApJ, 529, 338
\bibitem{rochapinto98} Rocha-Pinto, H. J., \& 
Machiel, W. J. 1998, A\&A, 339, 791.
\bibitem{rwi}Rogers, J., Swenson, F.J., \& Iglesias, C. 1996, ApJ, 456, 902
\bibitem{san}Sandquist, E., Taam, R., Lin, D. N. C., \& Burkert, A.
1998, ApJ, 506, L65
\bibitem{santos2000} Santos, N. C., 
Israelian, G., \& Mayor, M. 2000, A\&A, 363 228
\bibitem{s95}Saumon, D., Chabrier, G. \& Van Horn, 
H.M. 1995, ApJSupp, 99, 713
\bibitem{s96}Sills, A., Pinsonneault, M.H., \& Terndrup, D. 
2000, ApJ, 534, 335
\bibitem{s97}Soderblom, D.R., Fedele, S. B., Jones, B. F.,
Stauffer, J. R., \& Prosser, C. F. 1993, AJ, 106, 1059
\bibitem{stah88}Stahler, S. W., 1988, ApJ, 332, 804
\bibitem{twarog80} Twarog, B. A. 1980, ApJ, 242, 242
\bibitem{wg95} Wyse, R. F. G., \& Gilmore, G. 1995
AJ, 110, 2771


\end{thebibliography}
\end{document}